\documentclass[a4paper,twocolumn,showpacs,prl]{revtex4}%
\usepackage[latin9]{inputenc}
\usepackage{amsmath}
\usepackage{amssymb}
\usepackage{graphicx}
\usepackage{comment}
\usepackage{mathrsfs}
\usepackage{amsfonts}
\usepackage{float}%
\providecommand{\U}[1]{\protect\rule{.1in}{.1in}}
\makeatletter

\DeclareRobustCommand{\greektext}{  \fontencoding{LGR}\selectfont \def \encodingdefault{LGR}}
\DeclareRobustCommand{\textgreek}[1]{\leavevmode{\greektext #1}}
\DeclareFontEncoding{LGR}{}{}
\DeclareTextSymbol{\~}{LGR}{126}
\@ifundefined{textcolor}{}
{
\definecolor{BLACK}{gray}{0}
\definecolor{WHITE}{gray}{1}
\definecolor{RED}{rgb}{1,0,0}
\definecolor{GREEN}{rgb}{0,1,0}
\definecolor{BLUE}{rgb}{0,0,1}
\definecolor{CYAN}{cmyk}{1,0,0,0}
\definecolor{MAGENTA}{cmyk}{0,1,0,0}
\definecolor{YELLOW}{cmyk}{0,0,1,0}
}

\makeatother
\begin{document}
\title{Half metallic and insulating phases in BN/Graphene lateral heterostructures}
\author{Dong Zhang$^{1}$, Maosheng Miao$^{2}$, Fuhua Yang$^{1}$, Haiqing Lin$^{3}$ and Kai Chang$^{1,4}$}
\affiliation{$^{1}$SKLSM, Institute of Semiconductors, Chinese Academy of Sciences, P.O.
Box 912, Beijing 100083, China}
\affiliation{$^{2}$Materials Research Laboratory and Materials Department, University of
California, Santa Barbara, California 93106-5050, USA}
\affiliation{$^{3}$Beijing Computational Science Research Center, Beijing 100084, China}
\affiliation{$^{4}$CAS Center for excellence in quantum information and quantum physics}

\begin{abstract}
We investigate theoretically the electronic structure of graphene and boron
nitride (BN) lateral heterostructures, which were fabricated in recent
experiments. The first-principles density functional calculation demonstrates
that a huge intrinsic transverse electric field can be induced in the graphene
nanoribbon region, and depends sensitively on the edge configuration of the
lateral heterostructure. The polarized electric field originates from the
charge mismatch at the BN-graphene interfaces. This huge electric field can
open a significant bang gap in graphene nanoribbon, and lead to fully
spin-polarized edge states and induce half-metallic phase in the lateral
BN/Graphene/BN heterostructure with proper edge configurations.

\end{abstract}

\maketitle

Two-dimensional (2D) materials are promising to achieve the essential
requirements for future flexible\cite{flexible1,flexible2}, high speed and low
power-consumption electronic devices. Electrons in these 2D atomic crystals
such as graphene, hexagonal boron nitride (hBN), molybdenum disulphide
(MoS$_{2}$), other dichalcogenides and layered
oxides\cite{graphene1,graphene2,MoS21,MoS22,BN,2D}), behave like massless and
massive Dirac fermions\cite{grapheneES}, and display unique
transport\cite{grapheneTr1,grapheneTr2} and optical
properties\cite{grapheneOp}, such as Klein tunneling\cite{klein}, half-integer
quantum Hall effect at room temperature\cite{Hall1,Hall2} and valley-spin
locking\cite{valleyspin}. Recently, different isolated atomic crystals can be
assembled into designing heterostructures layer by layer via van del Waals
force between these 2D materials. These remarkable man-made new materials
reveal unusual properties and novel phenomena\cite{vdW1,vdW2,vdW3,vdW4,vdW5}.
Very recently, in-plane lateral heterostructures between graphene and
hexagonal boron nitride have been fabricated with controlled domain sizes and
shaped as combs, bars and rings\cite{lateral1,lateral2}. The lateral
heterostructure between graphene and hBN is particular important since
graphene is gapless semimetal whereas a monolayer of hBN is an insulator with
a wide bandgap of 5.9eV, and different atomic compositions may coexist within
continuous atomically thin films and many well-developed techniques have been
realized or attempted in graphene. The hybridized in-plane heterostructures
could provide us tremendous oppotunities towards atomically thin integrated
circuitry. With proper control the interface between graphene/hBN, the bandgap
and spin-relevant property could be precisely engineered, which are essential
parts in electronics.

Electronic transport in graphene is the subject of intense interest at
present. It might also be a promising material for spintronics and quantum
information and computation, owing to the low intrinsic spin-orbit
interaction, as well as the low hyperfine interaction of the electron spins
with the carbon
nuclei.\cite{spingra1,spingra2,spingra3,spingra4,spingra5,spingra6,QCgra,maggra1,maggra2,Allen}%
\ Recent theoretical and experimental results show that graphene could be the
long-awaited platform for spintronics, since graphene possesses the extremely
long spin diffusion length ($\sim$100$\mu m$) and high mobility at room
temperature. This could be a unique advantage for spintronic devices,
particularly for logic circuits in which information is coded by spin or
pseudo-spin (valley). It is interesting that despite there is no \textit{d}
band electrons, edge states in zigzag graphene nanoribbons (GNR) are spin
polarized with large magnetic interaction and ferromagnetically ordered at
each side. The spins at the two sides are oriented antiparallelly. This spin
distribution at the edges of GNR inspired an interesting idea that a strong
transverse electric field may drive the system into half metallic
phase,\cite{Louie} which is an intriguing class of materials, possessing
metallic phase for electrons with one spin orientation, but insulating for
electrons with the other. Fully spin-polarized electrical current in such
systems holds significant promise for spintronic devices. The spin property of
half-metal stimulates substantial efforts to search for half-metallic
materials containing d-shell electrons, for example, Heusler alloys and
manganese perovskite. However, in zigzag GNRs, one can relaize the
half-metallic phase under an in-plane electric field which is required very
high. For a GNR with 32 zigzag chains (32-GNR), one needs to apply 4.5
MV/cm$^{-1}$ electric field to achieve the half metallicity. The field needed
becomes larger for a narrower GNR. Applying such a strong transverse electric
field is challenging for the state-of-art gate technique; and so far, this
phenomenon has not been observed experimentally.

\begin{figure}[tbh]
\includegraphics[width=1\columnwidth]{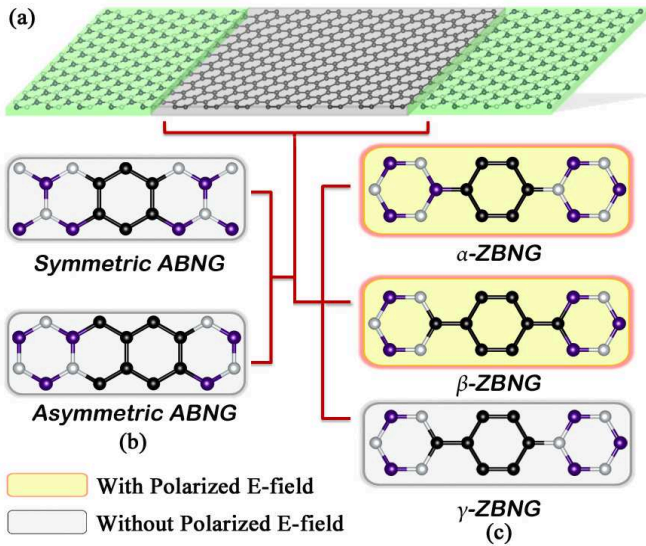}\caption{(color online) (a)
Schematic of the categories of lateral Boron Nitride-Graphene (BNG)
heterostructures. (b) The armchair-edged BNG (ABNG) heterostructures. The
higher panel shows symmetric ABNG (s-ABNG) while the lower one shows its
asymmetric counterpart (a-ABNG). (c) Three types of zigzag-edged BNG (ZBNG)
from top to bottom. The colors of each panel background indicate whether a
in-plane polarized electric field exsits or not. (d) Brillouin Zone folding of the BNG systems, the solid lines indicate the original BZ of graphene and BN, the dashed lines indicate the folded BZ.}%
\label{structure}%
\end{figure}

Here we propose a new route that require only the fabrication of materials.
Our design is inspired by the recent progress in fabricating in-plane
BN/graphene heterostructures using selective lithography
techniques.\cite{lateral1,lateral2} This technique can be used to grow
BN/Graphene/BN in-plane quantum well (QW) structures. The Graphene region
between the hexagonal BN is equivalent to a graphene nanoribbon, therefore we
denote this QW as BN/GNR/BN or BNG. We notice that depending on the
orientation of the QW, the interfaces may have equal number or unequal number
of B-C and N-C bonds. In the former case, the interface is charge neutral the
extra electrons filled in B-C bonding states are balanced by the excess
electrons from N-C bonds. However, if the interface has different counts for
B-C and N-C bonds, it will be charged. The charge will be balanced by the
other interface at the opposite side of GNR, and the whole QW is charge
neutral. The charges of different sign at the two interfaces will impose large
electric field to the GNR region, which can work as driving force toward half
metallicity as previously proposed.\cite{Louie}

We use first principles density functional calculations to demonstrate that
graphene in between BN behaves like GNR that can have strongly spin polarized
edge states, and more importantly, in some specific designs, may become
half-metallic. We use a slab model that contains a single layer of BNG and a
large vacuum region of about 20 Å~. Our calculations are based on the
Kohn-Sham formalism of density funcational theory (DFT) as implemented in the
Vienna Ab-initio Simulation Package (VASP)\cite{VASP}. The potentials of the
ions are represented by the projector augmented plane wave (PAW)\cite{PAW}
potentials. The generalized gradient approximation (GGA) in the framework of
Perdew-Burke-Ernzerhof (PBE) is adopted for the exchange-correlation
potential. All the first-principle calculations are performed using a
plane-wave cutoff of 600 eV on a $10\times10\times1$ Monkhorst-Pack k-point
mesh, and a 20 Å vacuum distance is used to ensure the fine decoupling between
adjacent slabs.

The possible BNG structures are illustrated in Fig. \ref{structure}. Similar
to GNR, the graphene ribbon sandwiched by BN can also be either armchair or
zigzag, denoting here as ABNG and ZBNG, respectively. As shown in Fig.
\ref{structure}, the ABNG can be further categorized into symmetric (s-ABNG)
and asymmetric (a-ABNG), depending on the exhibition of a mirror symmetry. The
interfaces in both structures are charge neutral, therefore no electric field
will be built up in them. Because zigzag GNR is the one that possesses
peculiar edge states, we will focus our study at ZBNGs. There are two major
configurations for ZBNG, distinguished by the fact that the BN and GNR atoms
at the interface are connected by one bond ($\alpha$-ZBNG) or by two bonds
($\beta$-ZBNG). While comparing the total energy, we find that $\alpha$-ZBNG
is more stable; its energy is 93.57 meV per atom lower than that of $\beta
$-ZBNG. In both cases, the interfaces are either B-C or N-C and therefore are
charged. The charged interface will naturally impose an electric field in the
GNR region. In order to compare the ZBNG with large intrinsic polarization
field, we also construct a $\gamma$-ZBNG, in which the interfaces consist of
only N-C bonds. Therefore, the two interfaces are equally charged and no
electric field will be induced.

\begin{figure}[tbh]
\includegraphics[width=1\columnwidth]{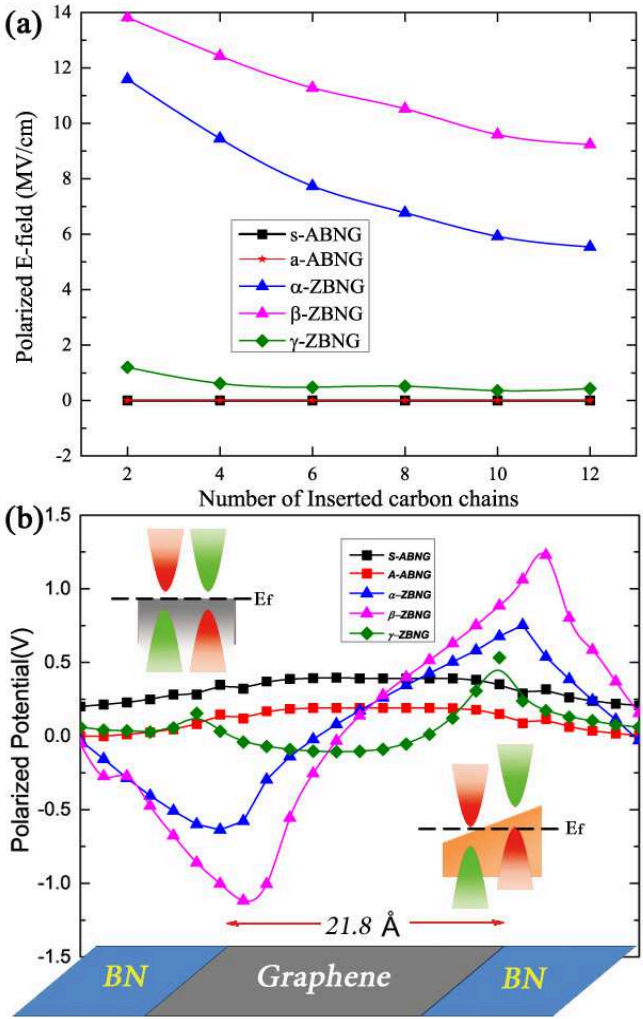}\caption{(color online) (a)
Estimated in-plane polarized electric field strength with different inserted
carbon chains i.e., the width of the GNR. (b) Schematic of in-plane polarized
potentials over the graphene region. The direction of voltage is perpendicular
to the BN/graphene interfaces and crossed the GNR regions. The width of GNR is
21.8 \r{A} (see the lowest inset). The two inserts display the behavior of
each spin under different type of in-plane potentials.}%
\label{field}%
\end{figure}

We first examine the strength of the intrinsic electric field in the BNG QWs.
Fig. \ref{field} presents the profiles of the electrostatic potentials in the
four different edge configurations of BNGs. The profiles for the s-ABNG and
a-ABNG are quite flat, consistent to the fact that there is no polarization
field as the interfaces of the two structures are neutral. The variations of
the profiles are from the small charge redistribution among the center and the
edge regions of BN and GNR. In sharp contrast to ABNG, both $\alpha$- and
$\beta$-ZBNGs exhibit very large potential change between the two BN-GNR
interfaces, which is a direct result of the polarization field from interface
charges. For a GNR region of 21.8 \r{A}~(including 12 zigzag or armchair C-C
chains), we found that the corresponding electric field are as large as 6.04
MV/cm and 10.24 MV/cm for $\alpha$- and $\beta$- ZBNGs, respectively. This is
much larger than the critical strength required for inducing half metallic
state in GNRs. Furthermore, there is no monotonic electric field in graphene
region of $\gamma$-ZBNG, which is expected. The large variation of the
potential profile is due to the large charge transfer from the interface
region to graphene and BN regions.

\begin{figure}[tbh]
\includegraphics[width=1\columnwidth]{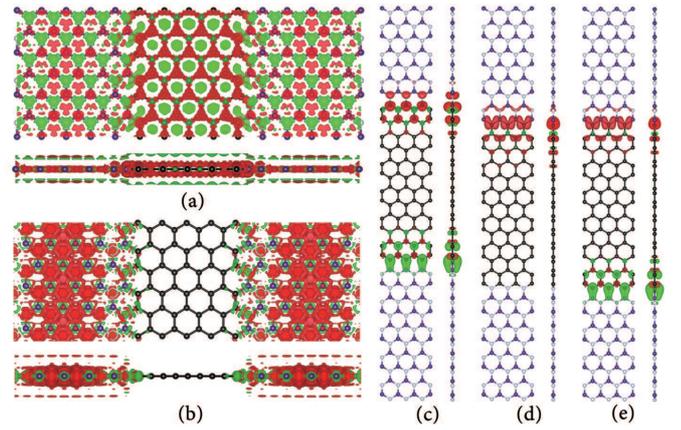}\caption{(color online) Schematic
of spin spatial distributions of each type of BNG (a) Spin spatial
distribution of s-ABNG. The upper and lower panels indicate the top-view and
side-view of the unitcell, respectively. (b) The same as (a), but for a-ABNG.
(c) Spin spatial distribution of \textgreek{a}-ZBNG. The left and right panels
indicates the top-view and side-view of the unitcell. (d) The same as (c), but
for \textgreek{b}-ZBNG. (e) The same as (c), but for \textgreek{g}-ZBNG. }%
\label{spin}%
\end{figure}

\begin{figure}[tbh]
\includegraphics[width=1\columnwidth]{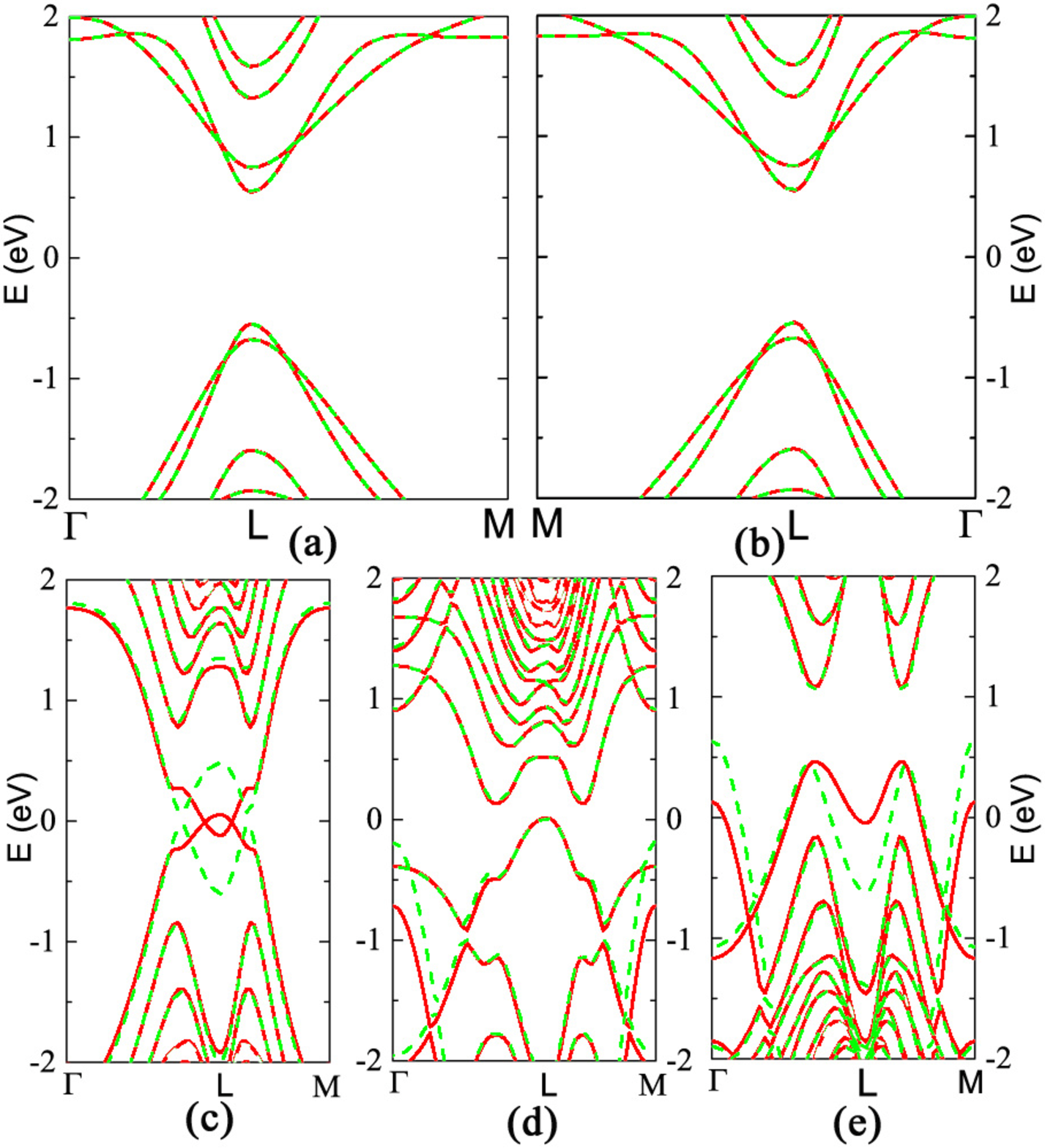}\caption{(color online) Band
structures and DOS of electron in the a-ABNG (a), s-ABNG (b), \textgreek{a}-ZBNG (c),
\textgreek{b}-ZBNG (d), and  \textgreek{g}-ZBNG (e). In each panel, the red
solid line indicates the spin-up branches, and the green dash line indicates
the spin-down branches.}%
\label{bands}%
\end{figure}

The top and side views of the spatial spin distributions in planar supercells
are illustrated in Fig. \ref{spin}. For ABNGs, the results are consistent with
the fact that there is no edge states for armchair GNRs. However, the spin
distributions are different for s- and a-ABNGs. For s-ABNG, the up-spin and
down-spin densities distributed uniformly through out the supercell; whereas
in a-ABNG, the GNR regions shows neither spin-up nor spin-down density,
indicating non-magnetic states. We also calculated the band structures and
found that both s-ABNG and a-ABNG exhibit large direct band gaps locating at
$\Gamma$ point [Fig. \ref{bands}(a) and (b)]. This is consistent to the
existence of sizable gaps in armchair GNRs. The calculations also show that
the bands of the two spin channels overlap with each other. Hexagonal BN has a
much larger gap than GNR, and the band edge states are in the gap of BN.
Therefore, both s-ABNG and a-ABNG are nonmagnetic and forms type one QW.

In contrast to ABNGs, the spin densities are highly localized around the
interfaces for both $\alpha$- and $\beta-$ZBNGs. This is due to the fact that
the spin polarization mainly happens to the edge states in the zigzag GNR
region in ZBNG. For $\alpha$-ZBNG, the spins oriented in opposite directions
at the two opposite interfaces. The band structure [Fig. \ref{bands}(c)]
further reveals that the $\alpha$-ZBNG is half metallic, since its spin-up
channel is metallic and spin-down channel is semiconducting with a gap of
0.193 eV. The mechanism of this induced half metallicity is the same to the
half metallicity in zigzag GNR, except that in ZBNG, the electric field is
intrinsically induced by charge accumulation. The exceedingly large electric
field from the interface charges allows the realization of half metallicity
for a very narrow graphene region in ZBNG. While assuming that the
accumulation of opposite charge on each interface does not change with the
width of graphene region, we can estimate that $\alpha$-ZBNG containing up to 16 number
of C-C zigzag chains (29.2 \r{A}~ of graphene region) or $\beta-$ZBNG containing up to 32 number
of C-C zigzag chains (60 \r{A}~ of graphene region) should be already be
half metallic.

For $\beta$-ZBNG, we found that spin density only localized at the right
interface and is spin up, which is distinctively different to $\alpha$-ZBNG.
Noticing that the polarization field is larger in $\beta$-ZBNG than in
$\alpha$-ZBNG, we think the reason os that the strong electric field induces
large charge transfer from left to the right side and the edge state at the
left side is no long occupied. $\gamma$-ZBNG show large spin distribution at
both left and right sides and with opposite orientations. However, the band
structure shows that it remains semiconducting in both spin-up and spin-down
channels (see Fig. 3(e)).

In conclusion, we propose an approach to achieve two dimensional half-metallic
systems based on graphene and hexagonal BN lateral heterostructures. We
demonstrate that the large polarization field originated from the interface
charge accumulation can drive the zigzag graphene ribbons sandwiched between
hBN into half metallic state in which the two spin channel locate at the
opposite interfaces. The advantage of the intrinsic electric field is the
avoidance of applying exceedingly large electric field at nanoscale.
Considering the recent progress in fabricating BN-Graphene heterstructures, it
is reasonable to expect the half metallic graphene can be observed
experimentally using our proposed design.

\end{document}